%% file: main.tex
\documentclass[10pt,twocolumn]{article} % Needs to be here for Overleaf
\input{header.tex}

\begin{document}

\twocolumn[
  \begin{center}
    \large
    \textbf{Stable Machine Learning Potentials for Liquid Metals via Dataset Engineering}
  \end{center}
  Alex Tai$^1$, Jason Ogbebor$^1$,
  Rodrigo Freitas$^1${\footnotemark[2]} \\
  $^1$\textit{\small Department of Materials Science and Engineering, Massachusetts Institute of Technology, Cambridge, MA, USA} \\

  {\small Dated: \today}
  
  \vspace{-0.15cm}
  \begin{center}
    \textbf{Abstract}
  \end{center}
  \vspace{-0.35cm}
  Liquid metals are central to energy-storage and nuclear technologies, yet quantitative knowledge of their thermophysical properties remains limited. While atomistic simulations offer a route to computing liquid properties directly from atomic motion, the most accurate approach—\textit{ab initio} molecular dynamics (AIMD)—is computationally costly and restricted to short time and length scales. Machine learning interatomic potentials (MLPs) offer AIMD accuracy at far lower cost, but their application to liquids is limited by training datasets that inadequately sample atomic configurations, leading to unphysical force predictions and unstable trajectories. Here we introduce a physically motivated dataset-engineering strategy that constructs liquidlike training data synthetically rather than relying on AIMD configurations. The method exploits the established icosahedral short-range order of metallic liquids—twelvefold, near--close-packed local coordination—and generates ``synthetic-liquid'' structures by systematic perturbation of crystalline references. MLPs trained on these datasets close the sampling gaps that lead to unphysical predictions, remain numerically stable across temperatures, and reproduce experimental liquid densities, diffusivities, and melting temperatures for multiple elemental metals. The framework links atomic-scale sampling to long-term MD stability and provides a practical route to predictive modeling of liquid-phase thermophysical behavior beyond the limits of direct AIMD.  \vspace{0.4cm}
]
{
  \footnotetext[2]{Corresponding author (\texttt{rodrigof@mit.edu}).}
}

\section{Introduction}

Liquid metal properties play a central role across technologies spanning manufacturing, energy storage, and nuclear systems, including additive manufacturing, liquid-metal batteries, and reactor cooling. More specifically, liquid-metal density controls flow, buoyancy, and heat transport, diffusivity governs mass transport and phase-transformation kinetics, and melting temperature defines the operational window for processes involving the solid--liquid phase change. Despite this importance, liquid metal properties remain poorly characterized experimentally. While benchmark density data are available for many pure elemental melts, systematic measurements of alloy melt densities as a function of composition are scarce. Diffusivity is even less well constrained, with reliable experimental data limited or unavailable even for many single-element liquid metals. This scarcity reflects substantial experimental challenges, including extreme temperatures, chemical reactivity with containers, and gravity-driven convection, which often require containerless techniques such as electrostatic or magnetic levitation, microgravity experiments aboard the International Space Station, and access to large-scale facilities such as synchrotrons and neutron sources\autocite{tamaru2018status}.

An alternative to direct experimentation is to compute liquid-metal properties from atomic motion using molecular dynamics (MD) simulations. In MD, atomic trajectories are obtained by integrating interatomic forces, making the accuracy of the method directly dependent on how these forces are computed. The most accurate MD approach evaluates forces from first-principles electronic-structure calculations at each timestep and is therefore referred to as \textit{ab initio} molecular dynamics (AIMD). However, the high computational cost of these electronic calculations severely limits the accessible time and length scales, typically to systems containing a few hundred atoms and simulation times of a few picoseconds. Accurately capturing liquid behavior requires substantially larger system sizes and longer trajectories, with tens of thousands of atoms needed to mitigate finite-size effects and hundreds of picoseconds to nanoseconds required for convergence of dynamic properties such as diffusion. These combined spatial and temporal demands render AIMD prohibitively expensive for calculating most liquid properties.

To extend the accessible time and length scales of MD simulations, machine learning interatomic potentials (MLPs) have emerged as computationally efficient alternatives to first-principles calculations. In MLP-based MD, a machine learning model is trained on first-principles data, typically using configurations sampled from AIMD trajectories. Once trained, the model replaces explicit electronic-structure calculations, enabling simulations at substantially larger spatial and temporal scales while retaining an accurate description of interatomic interactions. Despite these advances, two major challenges limit the application of MLPs to liquids. First, MLP-driven MD exhibits instabilities, in which undersampled regions of configuration space lead to unphysical force predictions and unexpected trajectory failure. Second, validation of MLP predictions often relies primarily on test-set errors evaluated on data closely related to the training set.

In this work, we address these two challenges in  MLP-based simulation of liquid metals, enabling accurate prediction of key structural, kinetic, and thermodynamic properties in good agreement with experimental data across a range of systems. Our approach is physically motivated by the well-established icosahedral short-range order observed in metallic liquids\autocite{frank1952supercooling,schenk2002icosahedral,kelton2003first,lee2004difference}, in which atoms form densely packed twelvefold coordination environments akin to those in face-centered cubic crystals. By systematically perturbing crystalline structures to reproduce this local disorder --- while avoiding the use of AIMD to generate training data ---  we construct synthetic-liquid datasets that capture the essential short-range physics of the liquid state. These datasets enable stable and accurate MD trajectories across temperatures for a range of single-element metal systems, enabling direct validation of structural, dynamical, and thermodynamic liquid properties against experiment.

\section{Results}

\subsection{Limitations of AIMD-based training sets}

\begin{figure*}[!htb]
  \centering
  \includegraphics[width=\textwidth]{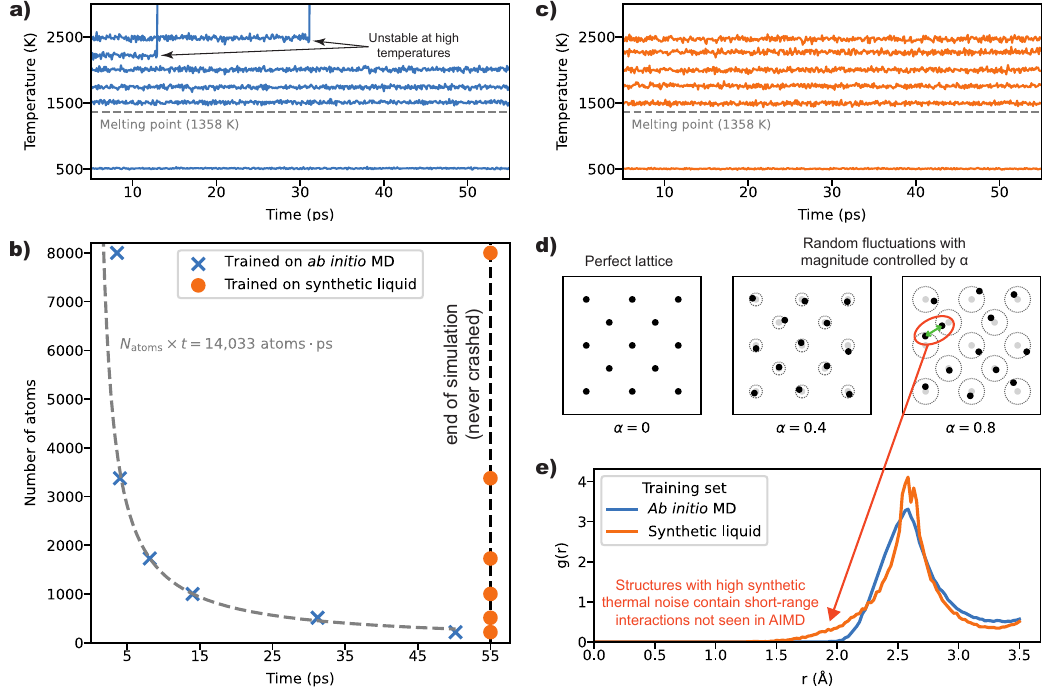}
  \caption{\label{fig:1} \textbf{Stabilizing liquid-phase MD with synthetic-liquid training.} \textbf{a)} Evolution of simulations driven by an MLP trained only on AIMD structures. Simulations destabilize and crash at high temperatures due to unphysical predictions. \textbf{b)} Simulation stability as a function of system size and trajectory length at 2500\,K. AIMD-trained MLPs destabilize earlier as system size increases, while SL-trained MLPs remain stable for all sizes. \textbf{c)} Temperature evolution for synthetic liquid MLPs showing stable trajectories across all temperatures. \textbf{d)} Synthetic liquid structures for training are generated by randomly perturbing ideal fcc lattice positions, with the magnitude of perturbation controlled by $\alpha$. This approach is physically motivated by the icosahedral order in metallic liquids. \textbf{e)} Radial distribution functions for the \textit{ab initio} and synthetic liquid training sets. The synthetic liquid set includes short-range interactions not present in AIMD.}
\end{figure*}

We select Cu as a representative system for testing MLP training methods and for systematically examining training-set construction and the limitations of AIMD-based approaches. We start by training an Atomic Cluster Expansion (ACE) MLP \autocite{drautz_atomic_2019, lysogorskiy_performant_2021, bochkarev2022efficient, lysogorskiy2023active} on liquid Cu AIMD snapshots to illustrate the common problem of MD instability \autocite{fu_forces_2023, stocker_how_2022, kandy_comparing_2023, gomes-filho_size_2023, kovacs_linear_2021, cao2025capturing, sheriff2024quantifying}. Figure~\rref{fig:1}{a} shows the energy over time for MLP-based MD simulations driven by an MLP trained on snapshots from AIMD runs. For the solid and the liquid near the melting point (1358\,K), the simulations are stable. However, at temperatures above the melting point, trajectories become unstable partway through the simulation, effectively crashing the MD simulation. This occurs despite the inclusion in the dataset of structures from AIMD simulations at temperatures up to 2500\,K.

The effect of increasing system size on MD stability is shown in fig.~\rref{fig:1}{b}, where it can be seen that all simulations using the AIMD-based MLP eventually destabilized. Failure occurs earlier in larger systems, with the product of the number of atoms and the time to failure remaining roughly constant. This behavior can be understood by considering stability in terms of atomic collisions: if any given collision has a small probability of producing an unstable configuration, the overall stability is determined statistically by the expected number of collisions before failure. In an equilibrium system, the total number of collisions scales linearly with both the number of atoms and the simulation time, leading to the observed size- and time-dependent destabilization. Such instability at larger time and length scales poses a challenge for calculating liquid properties of interest with MD, which all require tens of thousands of atoms and long trajectories to achieve proper convergence and statistical averaging.

\subsection{Synthetic liquid training sets}

To address this instability problem, we developed a ``synthetic liquid'' (SL) approach for constructing the training dataset for MLPs, illustrated in fig.~\rref{fig:1}{d}. This approach is physically motivated by the well-established observation that metallic liquids exhibit pronounced icosahedral short-range order\autocite{frank1952supercooling,schenk2002icosahedral,kelton2003first,lee2004difference}. Diffraction studies and simulations on undercooled Al, Ni, Fe, and other transition metals have shown that atoms in the liquid are not randomly arranged but instead form densely packed cages with twelve nearest neighbors, closely resembling the first coordination shell of a face-centered cubic (fcc) crystal (i.e., a cuboctahedron). The fcc cuboctahedron and the icosahedral cluster share nearly identical nearest-neighbor distances, differing only slightly in their angular topology. This geometric kinship provides a natural bridge between crystalline and liquid configurations: perturbing an fcc lattice generates radial and angular correlations characteristic of liquid metals while avoiding the need for costly AIMD sampling. Thus, by systematically introducing random displacements around fcc lattice sites, the SL method mimics the structural statistics of the liquid (i.e., dense local coordination and bond-angle dispersion) within a simple, controllable framework.

In the SL approach, thermal noise is mimicked by randomly displacing atoms from their ideal fcc lattice positions within a spherical volume. The radius of this volume is controlled by a parameter $\alpha$, which represents the magnitude of the imposed thermal noise and therefore plays a role analogous to temperature. When $\alpha = 1$, the sphere extends halfway to the nearest neighbor, i.e., the maximum possible size without overlapping volumes. The datasets consist of structures with a range of $\alpha$ values, uniformly distributed from 0 up to $\alpha_\t{max}$. Structures are also scaled up in size by applying uniform volumetric (isotropic) strains, with the magnitude uniformly varied from 0\% to 5\% relative to the ground-state structure. The parameter $\alpha$ can be directly related to the Lindemann melting criterion \autocite{lindemann1910calculation}, a well-known rule of thumb for determining when melting occurs:
\begin{equation}
    \sqrt{\left<\mu^2\right>} \ge \eta d,
\end{equation}
where $\mu$ is the instantaneous displacement of an atom from its lattice position, $\eta$ is the Lindemann coefficient, and $d$ is the nearest-neighbor distance. The brackets $\langle \cdot \rangle$ denote a thermal average over atomic vibrations, such that $\sqrt{\langle \mu^2 \rangle}$ represents the root-mean-square amplitude of atomic motion at a given temperature. Because $\alpha$ in the SL approach defines a comparable geometric displacement amplitude, it provides a convenient way to mimic the degree of thermal disorder, allowing a direct relationship between $\alpha$ and $\eta$ to be established:
\begin{equation}
    \alpha=\sqrt{\frac{20}{3}} \eta,
\end{equation}
see Supplementary Information sec.~1 for a derivation of this equation. Copper has $\eta = 0.108$ \autocite{vopson2020generalized}, which corresponds to $\alpha =0.279$. Thus, if $\alpha$ can be considered an effective temperature scale, then the ``melting point'' for SL Cu is $\alpha = 0.279$. We expect that datasets must include structures exceeding this value to capture liquid behavior.

Previous work has shown that MLP instability arises from unphysical model predictions resulting from insufficient coverage of the configuration space, particularly for structures with short interatomic distances\autocite{cao2025capturing, sheriff2024quantifying}. The SL approach provides a more comprehensive exploration of such environments. As shown in fig.~\rref{fig:1}{e}, the radial distribution function $g(r)$ of the SL dataset exhibits a longer left tail, indicating inclusion of short-distance interactions rarely observed in AIMD trajectories. Consequently, the SL-based MLP yields stable MD even at high temperatures, as shown in fig.~\rref{fig:1}{b} and \rref{fig:1}{c}. 

In summary, the SL approach can be viewed as a geometrically grounded method for reproducing the quasi-icosahedral coordination intrinsic to liquid metals through systematic perturbations of a crystalline reference. This dataset-engineering strategy bridges atomic-scale structure and simulation stability, ensuring that key short-range interactions governing liquid behavior are thoroughly represented in the training data.

\subsection{Optimizing parameters for synthetic-liquid training set}

We now examine how the choice of parameters in the SL approach influences MLP performance, focusing on the magnitude of the thermal noise $\alpha$. As shown in fig.~\rref{fig:1}{e}, short-range interactions are the main distinction between AIMD and SL training sets, making $\alpha$ a key parameter to investigate. To identify its optimal range, we construct multiple datasets with different maximum values $\alpha_\t{max}$. In each dataset, $\alpha$ is varied from 0 (the perfect crystal) up to $\alpha_\t{max}$. The largest value of $\alpha_\t{max}$ considered is 0.8; this limit was chosen because at $\alpha$ values approaching 1.0 the atoms become so close that numerical errors from the treatment of core electrons in density-functional theory (DFT) become significant\autocite{cryst9020086}.
\begin{figure*}[!htb]
  \centering
  \includegraphics[width=\textwidth]{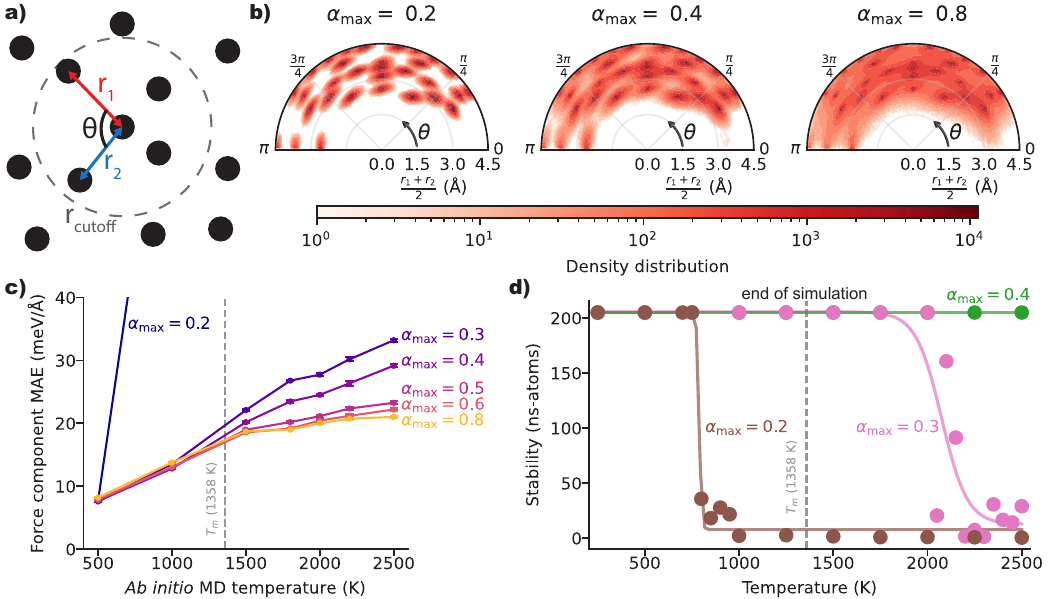}
  \caption{\label{fig:2} \textbf{Optimizing $\alpha_\t{max}$ for synthetic-liquid training.} \textbf{a)} Schematic of bonds pairs within cutoff radius defining the basis configuration space. \textbf{b)} Density distribution of local atomic environments for different $\alpha_\t{max}$ values. The radial component of the plot represents the average bond distance, while the angular component represents the bond angle. \textbf{c)} Mean absolute error of force predictions for MLPs trained on datasets with increasing $\alpha_\t{max}$, evaluated on independent AIMD test sets. Error bars represent the standard deviation of the (MAE) obtained from bootstrap resampling 100 times. \textbf{d)} Simulation stability (define as the product of simulation time and number of atoms) as function of temperatures for MLPs trained with different $\alpha_\t{max}$. Lines are included as a guide to the eye.}
\end{figure*}

To analyze how varying $\alpha_\t{max}$ influences the coverage of configuration space, we visualize the local atomic environments sampled by each MLP as a function of $\alpha_\t{max}$. Such environments can be represented using the distances and bond angles among atoms within the MLP cutoff radius $r_\t{cutoff}$, as shown in fig.~\rref{fig:2}{a}. The radial component of the plots in fig.~\rref{fig:2}{b} represents the average bond distance $(r_1 + r_2)/2$, while the angular component represents the bond angle for all interacting triplets of atoms. In the plot for $\alpha_\t{max}=0.2$, the distribution characteristic of the perfect crystal is clearly visible, with significant gaps in coverage remaining. As $\alpha_\t{max}$ increases, atomic deviations from ideal lattice positions progressively fill these gaps, and at high $\alpha_\t{max}$ values the remaining gaps correspond only to extremely short-distance structures. Analogous plots for AIMD snapshots are presented in Supplementary Information sec.~2, which reveal a similar lack of coverage at short interatomic distances and reinforce the conclusion drawn from fig.~\rref{fig:1}{e}: the SL approach systematically samples configurations that are rare or absent in AIMD trajectories, thereby eliminating undersampled regions of configuration space that give rise to MD instability.

To assess the effect of $\alpha_\t{max}$ on model accuracy, we evaluated the mean absolute prediction error of the force components using independent AIMD test sets. The results are shown in fig.~\rref{fig:2}{c}, with the mean absolute errors resolved by AIMD temperature. The analogous plot for energies is provided in the Supplementary Information sec.~5. When $\alpha_\t{max}=0.2$, the MLP exhibits very large errors except at the lowest temperature. Increasing $\alpha_\t{max}$ improves performance at high temperatures, but the marginal gains diminishing, with $\alpha_\t{max}=0.8$ showing no discernible improvement over $\alpha_\t{max}=0.7$ across the temperature range tested. See Supplementary Information sec.~3 for the effect of $\alpha_\t{max}$ directly on liquid properties.

As demonstrated in fig.~\rref{fig:1}{a}, stability depends on temperature. To explore the nature of the temperature-stability relationship, we quantify stability as the product of simulation time and number of atoms based on the inverse relationship found in fig.~\rref{fig:1}{b}. This stability is plotted against temperature for different $\alpha_\t{max}$ in fig.~\rref{fig:2}{d}. It is clear that for unstable potentials, there is a threshold temperature before which simulations can be reliably completed and beyond which they quickly destabilize. For $\alpha_\t{max}=0.2$ this temperature is around 800\,K ($0.59T_\t{m}$) and for $\alpha_\t{max}=0.3$ it is around 2000\,K ($1.47T_\t{m}$). All potentials with higher $\alpha_\t{max}$ were fully stable at the highest temperature simulated. 

\subsection{Property benchmarks and exchange--correlation dependence}
\begin{figure*}[!htb]
  \centering
  \includegraphics[width=\textwidth]{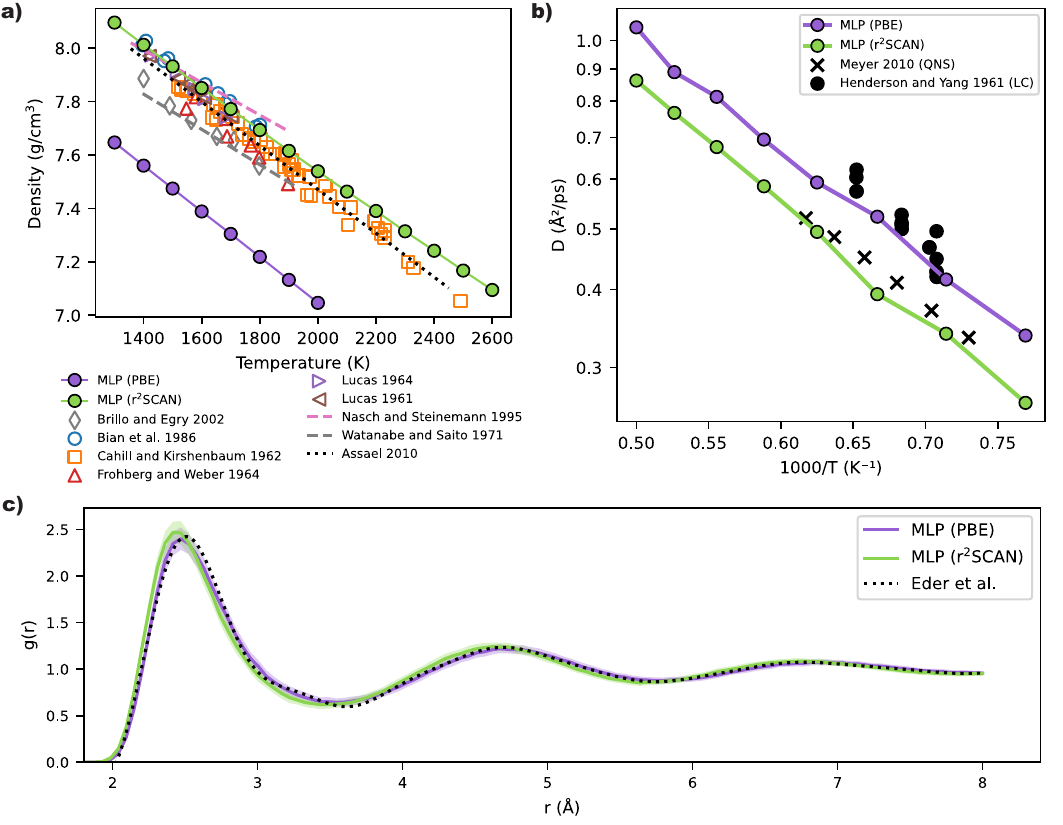}
  \caption{\label{fig:3} \textbf{Experimental validation and exchange--correlation functional effects in liquid Cu.} \textbf{a)} Temperature dependence of liquid density calculated from MLPs trained with PBE and r2SCAN exchange-correlation functionals, shown alongside experimental measurements\autocite{brillo_density_2003, cahill1962density, frohberg1964dichtemessungen, maoshu1986new, lucas1961densite, nasch_density_1995, watanabe1971densities, assael_reference_2010}, including a meta-analysis by Assael \textit{et al.}\autocite{assael_reference_2010}. 
  \textbf{b)} Temperature dependence of diffusivity of compared with experimental data from the long capillary method\autocite{henderson1961self} (filled circles) and quasi-elastic neutron scattering\autocite{meyer2010self} (crosses). \textbf{c)} Radial distribution function for liquid Cu at 1833\,K calculated using MLPs, in agreement with experimental results (shaded area is the standard error from the mean)\autocite{eder1980structure}.}
\end{figure*}
Having established that SL datasets with sufficiently high $\alpha_\t{max}$ consistently yield stable MLPs and lower errors, we now turn to the prediction of liquid Cu properties by the SL-trained MLPs and their comparison against available experimental data. Because these MLPs reproduce DFT forces and energies with high fidelity, their comparison with experiment primarily reflects the accuracy of the underlying DFT functionals, and we therefore also compare results obtained from MLPs trained on different exchange-correlation functionals. Figures~\rref{fig:3}{a} and \rref{fig:3}{b} show the calculated density and diffusivity of liquid Cu at different temperatures for MLPs trained on DFT data obtained using the PBE\autocite{perdew1996generalized} and r2SCAN\autocite{furness_accurate_2020} exchange-correlation functionals. PBE is a widely used generalized gradient approximation (GGA) functional, whereas r2SCAN is a more recent meta-GGA functional that systematically improves the description of bonding and equilibrium structures across a broad range of materials\autocite{tran2016rungs, zhang2018performance}. In our calculations (fig.~\rref{fig:3}{a}), the PBE-trained MLP underestimates the liquid density. The considerable scatter in the experimental data shown in the same figure highlights the inherent difficulty of measuring liquid-metal properties and suggests that MLPs trained with r2SCAN may already provide accuracy sufficient for practical applications --- a point revisited in sec.~\ref{sec:discussion}. The diffusivity results (fig.~\rref{fig:3}{b}) show a similar pattern of comparison, although interpretation is less straightforward because the two MLPs align with different experimental datasets. If precedence is given to the quasi-elastic neutron scattering measurements \autocite{meyer2010self} --- which directly probe atomic-scale diffusion and are typically performed under containerless conditions that eliminate convection and wall interactions --- the r2SCAN functional again provides closer agreement than PBE. Finally, fig.~\rref{fig:3}{c} compares the computed and experimental\autocite{eder1980structure} radial distribution function (RDF) of liquid Cu at 1833\,K, showing close agreement and confirming that the MLPs successfully reproduce the liquid structure.

\begin{figure*}[!htb]
  \centering
  \includegraphics[width=\textwidth]{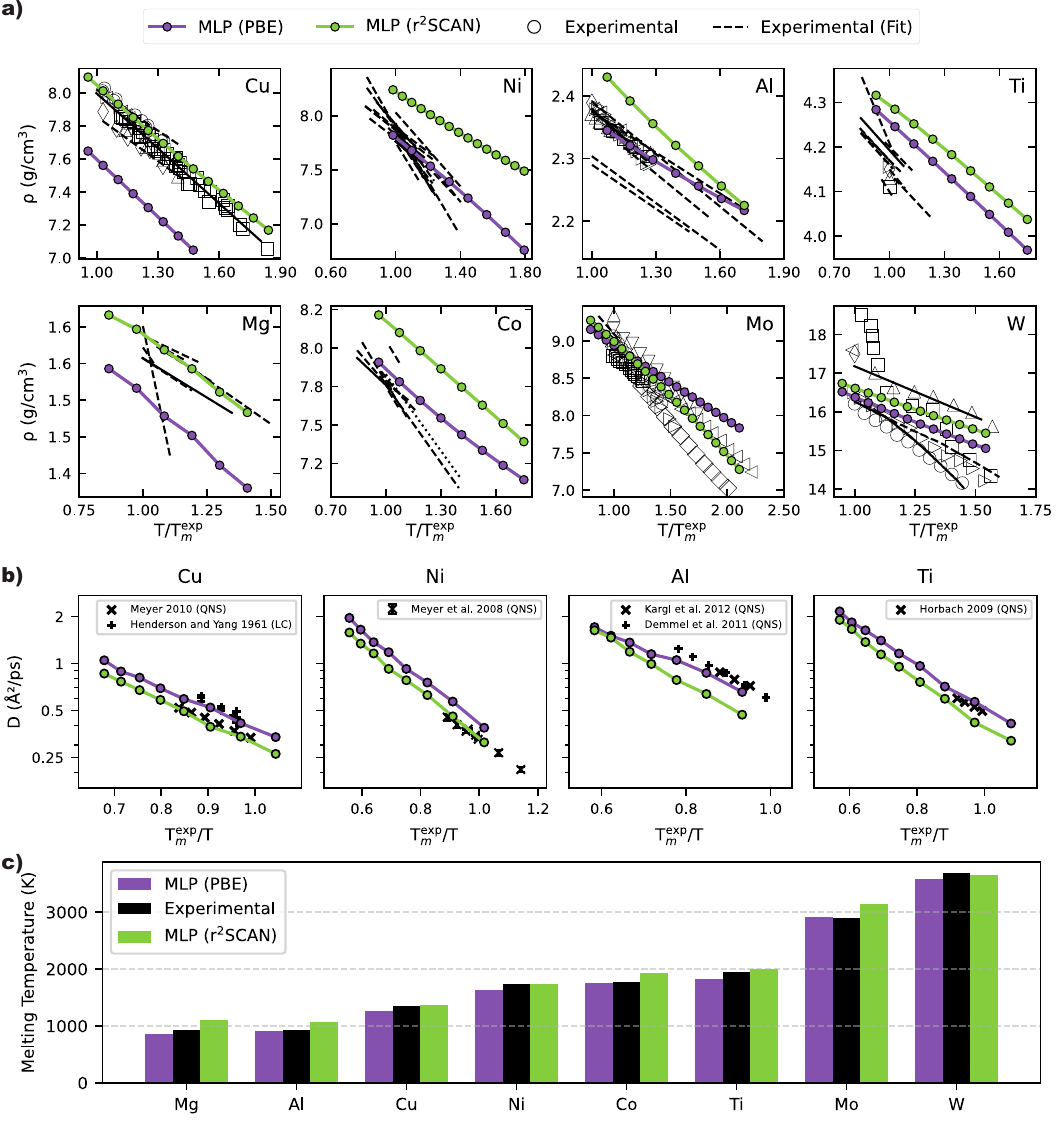}
  \caption{\label{fig:4} \textbf{Predicted liquid properties and melting points across elemental metals.}
  \textbf{a)} Temperature-dependent liquid density of Al \autocite{assael_reference_2006, smith_measurement_1999, yatsenko_notitle_1972, nasch_density_1995, levin_notitle_1968, coy_notitle_1955, gebhardt_notitle_1955, leitner_thermophysical_2017, peng_structural_2015, drotning_thermal_1979, schmitz_density_2012}, Co \autocite{wang_density_2024, paradis2008thermophysical, lee_measurement_2015, saito_density_1969, han_thermophysical_2002, brillo_density_2006, watanabe_thermophysical_2018, watanabe_densities_1971, mills2002recommended, abdullaev_density_2021, assael_reference_2012}, Cu \autocite{brillo_density_2003, assael_reference_2010, nasch_density_1995, maoshu1986new, watanabe_densities_1971, frohberg1964dichtemessungen, lucas1964specific, cahill1962density, lucas1961densite}, Mg \autocite{abdullaev_density_2019, mcgonigal_liquid_1962, gebhardt1955eigenschaften, grothe_liquid_1937, stankus_temperature_1990, arndt_density_1927, edwards1923density, pelzel_density_1941}, Mo \autocite{jeon_precise_2022, paradis_noncontact_2002, allen1963surface, shaner1977thermophysical, seydel_thermal_1979, minakov_ab_2018}, Ni \autocite{brillo_density_2003, saito_density_1969, nasch_density_1995, wang_density_2024, mills2002recommended, abdullaev_density_2015, ukhov_surface_1968, ayushina1969density, shiraishi_density_1964, kirshenbaum1963densities, lucas1960densite, saito_densities_1970, tavadze_notitle_1965, fang_density_2006, popel1969temperature, drotning1982thermal, lucas1972density, eremenko1964surface, stankus_thermophysics_1985, chung_noncontact_1996, ishikawa_thermophysical_2004, yoo_uncertainty_2015, kobatake_density_2013, brillo_density_2019, watanabe_densities_2016, mills2002recommended}, Ti \autocite{ozawa2017precise, saito_density_1969, wang_density_2024, lee_crystalliquid_2013, ishikawa2005thermophysical, saito1969density, paradis2000non, seydel_thermal_1979, jeon_effect_2016, amore_excess_2013, brillo_density_2019, watanabe_density_2019}, and W \autocite{leitner2019density, koval1997investigation, hess1999determination, hupf_electrical_2008, hixson1990thermophysical, berthault1986high, seydel_thermal_1979, allen1963surface, calverley1957determination} predicted by PBE- and r2SCAN-trained MLPs, shown alongside experimental data. Experimental values are plotted in black, with open markers representing individual measurements and dashed lines denoting empirical fits to the experimental datasets. Because detailed labeling would overcrowd the figure, the full list of experimental sources and fit parameters is provided in the Supplementary Information sec.~6.
  \textbf{b)} Diffusivity of Cu \autocite{meyer2010self, henderson1961self}, Ni \autocite{meyer_determination_2008}, Al \autocite{kargl_self_2012, demmel_diffusion_2011}, and Ti \autocite{horbach_improvement_2009}. These four elements are the only systems for which experimental diffusivity data could be found.
  \textbf{c)} Melting points of Mg, Al, Cu, Ni, Co, Ti, Mo and W \autocite{Okamoto2016-tf}. A more detailed analysis of the melting-point errors for each system is provided in the Supplementary Information sec.~8.} 
\end{figure*}

To assess the generality of the synthetic-liquid training strategy beyond Cu, we trained MLPs for seven additional single-element metals (Al, Co, Mg, Mo, Ni, Ti, and W). For each system, the training dataset was generated following the same procedure established for Cu, with the fcc structure was perturbed using $\alpha_\t{max}=0.6$. This approach ensures consistent sampling of thermally disordered configurations across systems while maintaining comparable coverage of short-range interactions. For systems whose ground state differs from fcc (e.g., bcc for W and hcp for Ti), that structure was also included in the training set to provide adequate description of the solid phase. A complete description of all datasets is provided in Supplementary Information sec.~4. Using these MLPs, we computed the liquid density, diffusivity, and melting point for each metal. These results, together with extensive experimental data collected from the literature, are shown in fig.~\ref{fig:4}.

To place the comparisons in figs.~\rref{fig:4}{a} and \rref{fig:4}{b} in context, it is important to note that experimental data often show substantial scatter for density and diffusivity, which makes it difficult to identify clear systematic deviations between MLP predictions and experiment. Despite this variability, distinct and reproducible trends emerge across exchange--correlation functionals. MLPs trained on r2SCAN calculations predict higher densities and lower diffusivities than those trained on PBE calculations. However, neither functional provides a consistent overall advantage in predictive accuracy of density and diffusivity relative to experiment, suggesting that the optimal choice is system-dependent. For the melting point (fig.~\rref{fig:4}{c}), r2SCAN systematically predicts higher values than PBE. But the comparison with experimental data is much clearer: the absolute deviation from experimental measurements for PBE ranges from -117\,K to 35\,K, corresponding to a relative errors of only -6\% to 1\% (Supplementary Information sec.~8).

Because of the results discussed above, we sought a simple heuristic for selecting the most suitable exchange--correlation functional based on readily available or easily calculated DFT data. The trends of each functional in predicting the density of the ground-state crystal phase are well established: PBE tends to underbind (overestimating lattice parameters and thus underpredicting density), while r2SCAN slightly overbinds (leading to overpredicted densities)\autocite{tran2016rungs, zhang2018performance}. These behaviors are straightforward to quantify through static DFT calculations of the solid phase. However, analogous trends for liquids are far less well known, since direct DFT simulations of liquid metals are limited by the small system sizes and short timescales accessible to AIMD.

\begin{figure}
    \centering
    \includegraphics[width=\linewidth]{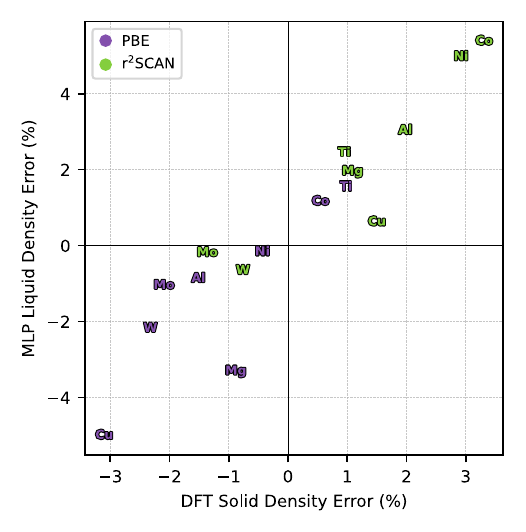}
    \caption{\label{fig:5} \textbf{Correlation between solid- and liquid-phase density errors.} Comparison of the relative error in liquid density at the melting point (from MLP simulations) with the corresponding solid density error obtained directly from DFT calculations.  The strong correlation between the two phases indicates that the well-known binding tendencies of each functional in the solid phase extend to the liquid phase as well.}
\end{figure}

To examine whether these solid-state trends extend to the liquid phase, fig.~\ref{fig:5} compares the liquid-density error at the melting point—obtained from MLP predictions—with the solid-density error computed directly from DFT relaxations. The errors were evaluated against experimental data: experimental liquid densities were taken from a single representative study for each element, prioritizing the most recent measurements, while experimental solid densities were corrected to 0\,K to enable comparison with DFT results (following the approach by Lejaeghere \textit{et al.}\autocite{lejaeghere2014error}). The correlation observed in fig.~\ref{fig:5} indicates that the systematic trends known for the solid phase extend to the liquid phase as well: functionals that overbind or underbind the solid exhibit the same tendency for the liquid. This correspondence implies that the most suitable exchange--correlation functional for liquid-phase simulations can often be inferred directly from the well-known solid-phase behavior.

\section{Discussion\label{sec:discussion}}

MLPs have rapidly expanded the reach of atomistic simulations, but their application to liquids remains limited by difficulties in both dataset generation and model validation. Our results show that these challenges can be systematically addressed by constructing training datasets that intentionally reproduce the statistical features of liquid disorder without relying on AIMD snapshots.

\subsection{Training sets and MD stability}

The high correlation among structures in AIMD trajectories restricts the sampled configuration space and limits model generality\autocite{drautz_atomic_2019}. While AIMD data remain the standard for liquid MLPs\autocite{liu_discrepancies_2023, balyakin_deep_2020, fang_molecular_2024, zeni_data-driven_2021, gomes-filho_size_2023, owen_complexity_2023, sivaraman_machine-learned_2020, zuo_performance_2020}, they poorly represent short-range atomic encounters that dominate liquid behavior. An obvious alternative—drawing configurations from purely random atomic arrangements—is not practical for metallic liquids, because such structures fail to reflect the known icosahedral short-range order. The SL approach addresses this by constructing training configurations through controlled random perturbations of the fcc crystal lattice (fig.~\rref{fig:1}{d}), which systematically include short interatomic distances absent in AIMD (fig.~\rref{fig:1}{e}). This construction is not \textit{ad hoc}: it is a compact surrogate for the icosahedral local structure of metallic liquids, ensuring that short-range repulsion and realistic bond-angle variability are both represented in training. Consistent with prior experimental and simulation studies showing quasi-icosahedral short-range order in metallic liquids\autocite{frank1952supercooling, schenk2002icosahedral, kelton2003first, lee2004difference}, the SL-trained MLPs also exhibit a finite population of icosahedrally coordinated atoms in the liquid phase (see Supplementary Information, sec.~9), illustrating that the SL dataset construction emphasizes physically relevant local environments.

This expanded coverage resolves one of the central problems in liquid-phase simulations: MD instability. AIMD-trained MLPs often destabilize at high temperatures due to unphysical predictions in sparsely sampled regions of the potential energy surface (fig.~\rref{fig:1}{a}). In contrast, SL-trained MLPs remain stable across all temperatures and system sizes (fig.~\rref{fig:1}{b} and \rref{fig:1}{c}). We find that the onset of instability scales with the product of the number of atoms and simulation time (fig.~\rref{fig:1}{b}), which reflects the total number of atomic collisions prior to failure. This statistical scaling indicates that stability is controlled by the likelihood of sampling underrepresented configurations during atomic interactions. Increasing $\alpha_\t{max}$ expands the explored region of configuration space and eliminates these poorly sampled regions, thereby improving robustness. The resulting MLPs sustain trajectories beyond the length and time scales required for computing liquid properties, establishing a direct connection between dataset engineering and long-term MD stability.

\subsection{Rigorous MLP validation}

Validation of MLPs has traditionally focused on errors in energies and forces on test sets drawn from the same data pool. However, low test-set error alone does not ensure physical reliability. For example, the model trained with $\alpha_\t{max}=0.3$ has force errors similar to those of larger $\alpha_\t{max}$ values at low temperature (fig.~\rref{fig:2}{c}) but rapidly destabilizes at high temperature (fig.~\rref{fig:2}{d}), illustrating that accurate interpolation does not guarantee robust extrapolation to configurations encountered during MD.

Validation through material properties provides a more stringent and physically meaningful assessment\autocite{cao2025capturing, sheriff2024quantifying}: RDFs reproduce the experimental liquid structure (fig.~\rref{fig:3}{c}), and property-level checks on density and diffusivity confirm accuracy over a broad range of conditions (figs.~\rref{fig:3}{a}, \rref{fig:3}{b}, \rref{fig:4}{a}, and \rref{fig:4}{b}). Experimental measurements themselves exhibit substantial scatter, often comparable to the full spread between independent studies, which limits the utility of pointwise comparisons between simulations and individual datasets. Within this context, the present results fall consistently within the experimental envelope, indicating that the approach achieves a level of fidelity comparable to the reproducibility of existing experimental measurements and is therefore suitable for practical calculations of liquid-metal properties where experimental data are sparse or inconsistent.

\subsection{Exchange--correlation dependence}

The accuracy of MLPs is ultimately bounded by the density-functional theory (DFT) functional used to generate the reference data. As shown in fig.~\ref{fig:3} and fig.~\ref{fig:4}, r2SCAN-based MLPs predict higher densities and lower diffusivities than their PBE counterparts, consistent with the overbinding tendency of r2SCAN in solids\autocite{tran2016rungs, zhang2018performance}. Figure~\ref{fig:5} confirms that this bias extends systematically from the solid to the liquid phase: functionals that overbind in the solid tend to yield denser liquids, while those that underbind produce lower liquid densities. This correlation provides a simple heuristic for selecting an appropriate exchange--correlation functional for liquid-phase simulations. While r2SCAN offers improved accuracy for most systems, its poorer description of magnetic metals such as Ni and Co\autocite{furness_accurate_2020} indicates that functional performance remains system-dependent.

\section{Conclusion}

Accurate and stable MLPs for liquid metals can be obtained without AIMD by constructing synthetic-liquid datasets that deliberately sample the relevant configuration space. This framework overcomes a long-standing practical obstacle: naive random placement of atoms produces abundant near-overlaps that neither reflect the known icosahedral short-range order of metallic liquids nor are compatible with standard pseudopotential DFT, which fails when core regions overlap—forcing prior work to rely almost exclusively on costly AIMD snapshots for training. Our method introduces this missing short-range order explicitly, using controlled perturbations of close-packed crystalline environments to reproduce the dense, twelvefold local coordination characteristic of metallic liquids while respecting excluded-volume constraints. By tuning the magnitude of atomic perturbations through $\alpha_\t{max}$ captures the short-range interactions that govern stability at high temperatures, enabling MLPs that reproduce experimental liquid densities, diffusivities, and melting points while maintaining numerical stability in large-scale MD simulations that are inaccessible to computationally costly AIMD.

Beyond accuracy, the present framework establishes a transparent and reproducible connection between configurational sampling and long-term MD stability, providing a physically grounded criterion for dataset construction and model validation. Although experimental measurements of liquid-metal properties exhibit substantial scatter across independent studies, the present calculations fall consistently within this experimental envelope. In this regime, agreement at the level of experimental variability represents the strongest validation achievable and underscores the practical value of the approach, particularly where experimental data are sparse, inconsistent, or difficult to obtain.

The predictive ceiling remains set by the underlying DFT functional; however, the correlation between solid- and liquid-phase density errors identified in fig.~\ref{fig:5} offers a practical heuristic for functional selection based on readily accessible solid-state calculations. Taken together, this dataset-engineering strategy provides a scalable route to predictive modeling of liquid-phase thermophysical behavior at a consistent level of physical theory, without resorting to AIMD. Because the SL approach is not tied to a specific element or crystal structure, it is readily extendable to multicomponent systems \autocite{sheriff2025machine}, where experimental data on liquid properties are virtually nonexistent. Its application to alloys enables systematic evaluation of composition-dependent thermophysical behavior, thereby advancing the computational design of materials and processes governed by liquid-metal properties.
\clearpage

\section{Methods}

\subsection{Density-functional theory}
\label{methods:DFT}

All density-functional theory (DFT) calculations were performed using the Vienna \textit{ab initio} Simulation Package (VASP)\autocite{kresse1999ultrasoft} version 6.2.1. The projector augmented-wave (PAW) method was employed with pseudopotentials from the \texttt{potpaw.64} library, and the electronic exchange--correlation energy was described using the generalized-gradient approximation (GGA) in the Perdew--Burke--Ernzerhof (PBE) formulation\autocite{Kresse1993-yi, Kresse1996-lx, Kresse1996-rj, Kresse1994-qo, kresse1999ultrasoft} and the meta-GGA in the r2SCAN formulation\autocite{furness_accurate_2020}.

\begin{table}[h]
\centering
\begin{tabular}{|l|c|c|}
\hline
\textbf{Pseudopotential} & \makecell{\textbf{Energy}\\\textbf{cutoff (eV)}} & \makecell{\textbf{Minimum}\\\textbf{distance ($\Ang$)}} \\
\hline
Al & 400 & 45 \\
Cu\_pv & 550 & 40 \\
Cr\_pv & 400 & 45 \\
Mg\_pv & 500 & 45 \\
W\_sv & 350 & 45 \\
Mo\_pv & 400 & 55 \\
Co & 400 & 30 \\
Ni\_pv & 500 & 40 \\
\hline
\end{tabular}
\caption{Converged DFT parameters for each element to within 1\,meV/atom. Energy cutoffs and minimum distances are similar to the recommended values by VASP\autocite{blochl1994projector, kresse1999ultrasoft} and the KpLib database\autocite{wisesa2016efficient, wang2021rapid}, respectively.}
\label{tab:pseudopotentials}
\end{table}

Reciprocal-space integrations were performed using $k$-point grids generated with the KpLib algorithm\autocite{wisesa2016efficient,wang2021rapid}. The minimum allowed distance between lattice points in the real-space superlattice was set to the converged values listed in table~\ref{tab:pseudopotentials}. $\Gamma$-centered grids were excluded in all cases. Electronic occupancies were evaluated using Methfessel--Paxton smearing with a width of 0.2\,eV. Spin polarization was enabled only for calculations involving Co and Ni. Static self-consistent field (SCF) cycles were converged to a total energy tolerance of $10^{-6}$~eV.

\subsection{Ab initio molecular dynamics}
\label{methods:aimd}

AIMD simulations were performed for elemental Cu using a 108-atom face-centered cubic (fcc) supercell. All simulations were carried out in the isothermal--isobaric (NPT) ensemble at temperatures ranging from 100\,K to 2500\,K and zero external pressure. Temperature and pressure were controlled using a Langevin thermostat and Langevin barostat, each with a friction coefficient of $10~\t{ps}^{-1}$ applied to both atomic and lattice degrees of freedom. Ionic motion was integrated with a timestep of 3\,fs, and each trajectory was propagated for at least 9\,ps to ensure adequate equilibration at every temperature.

Electronic structure evaluations during AIMD employed only the $\Gamma$ point for Brillouin-zone sampling. Methfessel--Paxton smearing with a width of $\sigma = 0.1$~eV was used to determine electronic occupancies.

\subsection{Training set composition}
\label{methods:trainset}

Training datasets for the Cu SL MLPs consisted of 50 static DFT calculations on 108-atom supercells derived from the ground-state fcc crystal structure. Atomic positions were randomly displaced from their lattice sites within a spherical volume centered on the ideal lattice positions. The choice to perturb close-packed crystalline environments reflects established evidence that liquid metals organize into locally icosahedral, twelve-coordinated cages at nearly crystalline nearest-neighbor distances\autocite{schenk2002icosahedral,kelton2003first}. The displacement of each atom was parametrized in spherical coordinates as
\begin{equation}
    \phi = 2\pi x_{\phi}
\end{equation}
\begin{equation}
    \theta = \cos^{-1}(-1+2x_{\theta})
\end{equation}
\begin{equation}
    R = \alpha d \sqrt[3]{x_{R}}
\end{equation}
where $x$ is a random number uniformly distributed in the interval [0,1], $\alpha$ is a tunable displacement amplitude, and $d$ is half the nearest-neighbor distance in the ground-state crystal. To avoid overlapping atomic positions, $\alpha$ should not exceed 1.

Structures were generated with $\alpha$ values uniformly distributed from 0 up to $\alpha_\t{max}$, and with isotropic lattice scaling applied up to 5\% relative to the equilibrium lattice parameter. Following the optimization of $\alpha$ parameters, all datasets employed a $\alpha_\t{max}$ of 0.6.

We follow a similar procedure for generating SL datasets for Al, Cr, Mg, W, Mo, Co, and Ni. For systems whose ground state differs from fcc (e.g., bcc for W and hcp for Ti), that structure was also included in the training set to provide adequate description of the solid phase. Each dataset consists of 50 total structures, with 48 (fcc) or 52 (bcc, hcp) atoms per structure. A complete description of all datasets is provided in Supplementary Information sec.~4.

The interatomic MLP trained on AIMD-based data for Cu employed configurations extracted from the AIMD trajectories described in sec.~\ref{methods:aimd}. These AIMD simulations spanned temperatures from 100\,K to 2500\,K, covering both solid and liquid phases. Snapshots were sampled at intervals of at least 450\,fs, excluding the initial 1.5\,ps of equilibration. Each configuration was subsequently recomputed with static DFT calculations using the same computational parameters described in sec.~\ref{methods:DFT}, and the resulting data were aggregated into the AIMD-based training set.

\subsection{MLP training}
\label{methods:mliap}

ACE potentials were fit with pacemaker \autocite{drautz_atomic_2019, lysogorskiy_performant_2021, bochkarev_efficient_2022, lysogorskiy2023active}. The radial cutoff was set to $4.75\,\Ang$ for all potentials. The relative weighting factor between energy and force terms in the loss function, $\kappa$, was set to 0.1. A body-order ladder scheme was employed, incrementally increasing the number of basis functions by steps of 25. For each ladder step, the training was performed for up to 1400 iterations. The final MLP contained 224 basis functions and a total of 808 trainable parameters. To ensure physical behavior at very short interatomic distances, a Ziegler--Biersack--Littmark (ZBL) core repulsion potential was applied to all atom pairs at separations shorter than the minimum bond distance present in the corresponding dataset\autocite{ziegler1985stopping}.

\subsection{Molecular dynamics}

Molecular dynamics (MD) simulations with the MLPs were performed using the Large-scale Atomic/Molecular Massively Parallel Simulator (LAMMPS)\autocite{Thompson2022} to evaluate structural and transport properties, as well as to assess model stability across system sizes and temperatures. A table with summary of all computed properties is available in Supplementary Information sec.~7.

For the calculation of the RDF in figs.~\rref{fig:1}{e} and \rref{fig:3}{c}, simulations were initialized from 108 randomly positioned, non-overlapping atoms. Each trajectory was run for 200\,ps in the isothermal-isobaric (NPT) ensemble using a Nos\'e-Hoover thermostat relaxation time of 1\,ps and Nos\'e-Hoover barostat relaxation time of 10\,ps \autocite{martyna1994constant, tuckerman2006liouville, shinoda2004rapid, parrinello1981polymorphic}. The initial 5\,ps of equilibration were excluded from the RDF analysis.

To evaluate the MD stability of MLPs in figs.~\rref{fig:1}{a--c} and \rref{fig:2}{d}, simulations were initialized from random atomic positions and equilibrated for 5\,ps in the NPT ensemble (thermostat 0.1\,ps, barostat 1\,ps) at 2500\,K. The systems were then evolved for 50\,ps in the microcanonical (NVE) ensemble. A trajectory was classified as destabilized if any atoms were lost or if the instantaneous temperature exceeded 4000\,K.

For density calculations in figs.~\rref{fig:3}{a} and ~\rref{fig:4}{a}, systems containing 10,000 atoms were initialized from random positions and equilibrated for 60\,ps in the NPT ensemble with thermostat and barostat relaxation times of 1\,ps and 10\,ps, respectively. The density was obtained by averaging the simulation cell volume over the final 40\,ps of the trajectory.

For diffusivity calculations (figs.~\rref{fig:3}{b} and ~\rref{fig:4}{b}), systems of 10,000 atoms were first equilibrated for 20\,ps in the canonical (NVT) ensemble using a thermostat relaxation time of 1\,ps, with the simulation volume fixed to that obtained from the corresponding density calculation. The systems were then propagated for 40\,ps in the NVE ensemble, and the mean-squared displacement (MSD) of the atoms was tracked over time. The self-diffusion coefficient $D$ was computed from the linear regime of the MSD curve as
\begin{equation}
    D = \frac{\langle |\mathbf{r}(t) - \mathbf{r}(0)|^2 \rangle}{6t}
\end{equation}
where t is time and $\mathbf{r}$ is distance from the initial position such that the numerator of the right hand side is the MSD. The temperature dependence of diffusivity fitted to the Arrhenius relation:
\begin{equation}
    D=D_0 \exp\l(\frac{-\Delta E}{k_\t{B}T}\r),
\end{equation}
where $D_{0}$ is the pre-exponential factor, $\Delta E$ is the activation energy, $k_\t{B}$ is the Boltzmann constant, and $T$ is the temperature.

\subsection{Melting point calculation}

The melting temperatures of the MLPs (fig.~\rref{fig:4}{c}) was determined by calculating the free energies of the solid and liquid phases across their stable and metastable temperature ranges and identifying the temperature at which these phases are in thermodynamic equilibrium. To calculate the free-energies, we employed the non-equilibrium thermodynamic integration (NETI) approach\autocite{Jarzynski1997, Oberhofer2005, Freitas2016}, using LAMMPS to perform the necessary simulations. We first computed the free energies of both phases at the extremes of their selected temperature ranges, followed by the temperature dependence computed via the reversible scaling method\autocite{deKoning1999}. All simulations used a timestep of 1\,fs.

For solid phases, we used the Frenkel-Ladd method\autocite{Frenkel1984, Freitas2016}, in which we compute the free-energy difference between the MLP of interest and an idealized Einstein crystal. The spring constant and equilibrium volume were determined from simulations in the isothermal-isobaric ensemble, which ran for a total of 50,000 timesteps using systems of 2,000 atoms for bcc and 2,048 atoms for fcc. The subsequent constant-volume NETI simulation ran for 10,000 timesteps for equilibration and 50,000 timesteps for the non-equilibrium switching procedure with systems of 11,664 atoms for bcc and 10,976 atoms for fcc. The reversible scaling simulations used the same number of atoms as the Frenkel-Ladd simulations, and the same timing parameters for equilibration and switching. Previous work\autocite{Freitas2016} has demonstrated the effect of system size and timing parameters on the precision of these methods, and the selected sizes and duration are sufficient for a well-converged result.

For liquid phases, we used the Uhlenbeck-Ford (UF) model\autocite{Paula2017} as the reference state in our NETI simulations\autocite{liquid_free_energy}, for which the absolute free energy can be computed exactly with the help of pre-calculated virial coefficients for select values of the interaction parameters\autocite{Paula2016}. Simulations in the isothermal-isobaric ensemble followed the same details as described for the solid phases. For the subsequent NETI simulations, the parameters of the UF model were set to $p = 75$ and $\sigma = 1.5$ for all MLPs. We found that these parameters are a good choice for all liquid metals in general. An increase in $p$ and/or $\sigma$ did not result in any significant improvement in the precision or efficiency of the NETI method. Timing parameters for the reversible scaling simulations are the same as the ones used for the solid phases, and system sizes were also chosen to exactly match the corresponding solid phase.

\section{Data and code availability}

The interatomic MLPs described in this work, along with the DFT calculations used to fit them, will be deposited to the ColabFit Exchange via the OpenKIM framework. The models will be assigned permanent API-identifiers in the OpenKIM repository and made publicly accessible via ColabFit. Any additional scripts, input files, or analysis code associated with the project can be shared through a GitHub repository upon  request.

\section{Author contributions}

J.O. performed all melting-point calculations. A.T. carried out all the other simulations and all subsequent data analysis. All authors contributed to the interpretation of the results, drafted the manuscript, and approved the final version of the paper.

\section{Acknowledgments}

This work was supported by the Portuguese Foundation for International Cooperation in Science, Technology and Higher Education in the MIT--Portugal Program. The authors acknowledge the MIT Office of Research Computing and Data for providing high performance computing resources that have contributed to the research results reported within this paper.

\section{Competing interests}

The authors declare no competing interests.

\clearpage
\twocolumn[
  \begin{center}
    \Large
    \textbf{References}
  \end{center}
]
\printbibliography[heading=none]

\end{document}

%% file: header.tex
\usepackage{graphicx} % Command \includegraphics
\graphicspath{{figures}} % Directory to search for figures
\usepackage[font=small,labelfont=bf]{caption} % Caption control

\usepackage{xcolor}
\definecolor{black}{HTML}{212427}
\definecolor{blue}{HTML}{0563C1}
\definecolor{red}{HTML}{B51700}

\usepackage{hyperref,nameref}
\hypersetup{colorlinks=true,allcolors=blue}
\newcommand{\rref}[2]{\hyperref[#1]{\ref{#1}#2}} % Hyperref for fig. letters

\color{black}
\usepackage{setspace} % Line spacing
\usepackage[margin=0.67in]{geometry} % Margin size
\usepackage[htt]{hyphenat} % Allows \texttt to break
\usepackage[switch]{lineno} % Line numbers (for review)
\usepackage{makecell}

\usepackage{fancyhdr,lastpage}
\usepackage{titlesec}
\titleformat{\section}{\Large\bfseries}{\thesection.}{5pt}{}
\titleformat{\subsection}{\bfseries}{\thesubsection.}{3pt}{}
\titlespacing{\section}{0pt}{\baselineskip}{0pt}
\titlespacing*{\section}{0pt}{\baselineskip}{0pt}
\titlespacing*{\subsection}{0pt}{\baselineskip}{0pt}

\usepackage{amsmath,amssymb}
\newcommand{\Ang}[0]{\mathring{\mathrm{A}}} % Angstrom symbol
\renewcommand{\l}[0]{\left} % Shortcut for left modifier
\renewcommand{\r}[0]{\right} % Shortcut for right modifier
\renewcommand{\t}[1]{\text{#1}} % Text mode

\usepackage{multirow} % Multi row/column in tables
\usepackage[normalem]{ulem}

\usepackage[
  backend=biber,        % Use biber
  autocite=superscript, % \autocite results in superscript, \cite is on line
  sorting=none,         % References appear in the order they are cited
  style=numeric-comp,   % Cite using numbers and compact format (e.g., 1-3)
  maxnames=100,         % Number of authors that trigger et al.
  minnames=100,         % Number of authors before et al.
  giveninits=false,     % Only initials for first name
  autopunct=false,      % Put citation before punctuation marks
  doi=true,             % Show or not DOI number
  hyperref=true         % Hyperref links in the bibliography (mostly DOI)
]{biblatex}
\DeclareFieldFormat{labelnumberwidth}{\;[#1]\!\!\!\!} % Square brackets entry
\renewbibmacro{in:}{} % Remove "in: Journal" from reference
\AtNextBibliography{\small} % Bibliography font size
\AtEveryBibitem{\clearfield{pages}} % Don't show page
\AtEveryBibitem{\clearfield{volume}} % Don't show volume
\AtEveryBibitem{\clearfield{number}} % Don't show number
\AtEveryBibitem{\clearfield{month}} % Don't show month
\AtEveryBibitem{\clearfield{url}} % Don't show url
\AtEveryBibitem{\clearfield{urlyear}} % Don't show visited on
\AtEveryBibitem{\clearfield{urlmonth}} % Don't show visited on
\AtEveryBibitem{\clearfield{urlday}} % Don't show visited on